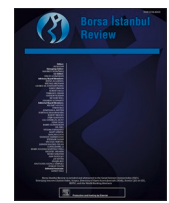

# Does the financialization of agricultural commodities impact food security? An empirical investigation

Manogna R. L.[*], Nishil Kulkarni

*Department of Economics and Finance, BITS Pilani K K Birla Goa Campus, India*

A B S T R A C T

The possible effect of the financialization of agricultural commodities on food security has become an evolving worry in recent years. This study seeks to empirically investigate this complicated problem and influence policy choices to ensure a more stable and secure food system by analyzing the role of financialization in global food markets. The study uses the panel data regression model, moderating effects model, and panel data regression with threshold variable to analyze financialization due to three agricultural commodities: wheat, maize, and soybean. For wheat, maize, and soybean futures traded on the Chicago Board of Trade, we utilize data related to annual trading volume, annual open interest contracts, and a ratio of annual trading volume to annual open interest contracts. The sample covers five developed countries - the United States, Australia, Canada, France, and Germany, and seven developing countries- China, Russia, India, Indonesia, Brazil, Vietnam, and Thailand. Annual panel data are constructed for the 2000–2021 period. The Human Development Index (HDI) is the threshold variable to differentiate the impact across these countries. The findings reveal that the financialization of agricultural commodities has negatively impacted food security globally, with wheat and soybean having a greater negative impact than corn. Also, there is a more considerable impact on developing countries compared to developed countries. The study finds that monetary policy can potentially reduce the impact of agricultural financialization on food security. The findings of this paper act as a guide to assist policymakers in ensuring that the world's food supply stays secure and available.

## 1. Introduction

Food security is a critical global problem, owing to the millions still starving and malnourished. A projected 811 million individuals will be hungry in 2020, up from 690 million in 2019. Furthermore, the COVID-19 pandemic has worsened food instability, with an extra 161 million people facing hunger in 2020 due to the pandemic's economic and societal disruptions (FAO, 2020; FAO, 2021). According to the U.N., the global population will hit 9.7 billion by 2050, resulting in a 60% rise in food consumption (FAO, 2017). Rapid population growth, climate change, and political upheaval have raised worries about the capacity of the world's agricultural system to satisfy everyone's requirements. Climate change impacts food production because variations in temperature and rainfall trends can result in reduced agricultural yields and supply chain disruptions. Furthermore, political insecurity and war disrupt food production and delivery, resulting in food scarcity and malnutrition. As a result, guaranteeing food security has become a top concern for governments, organizations, and people all over the globe. Fig. 1 shows variations in the food price index over the years.

In recent years, the financialization of agricultural goods has been connected to several high-profile food crises. The 2008 global food crisis, which resulted in sharp increases in food costs and widespread demonstrations and riots in many countries, is frequently mentioned as a critical illustration of the possible effect of financialization on food security. Financialization of agricultural products refers to speculating on the price of agricultural commodities using financial tools such as commodity futures and options contracts. In this context, speculators buy futures contracts for commodities like wheat, soybeans, or maize in the hope of benefiting from price rises in the future. Fig. 2 shows the fluctuations in commodity futures on corn, rice, soybean, and wheat since the deregulation in the U.S. and elsewhere.

While these tools can help farms and other market players control risk and secure consistent revenue, speculators can also use them to manipulate markets and increase price volatility. Financial instruments used in agricultural markets can provide liquidity, help in price discovery, and bring risks. The increased participation of financial speculators in farm markets has resulted in higher price fluctuations, raising the possibility of speculative bubbles and market volatility. Fig. 3 shows the price fluctuations of the four major staple grains in the international market from 1992 to 2019. While these tools can help farmers control risk, they may face challenges such as higher supply costs or reduced income when prices fall. As the global food system has become more interconnected and susceptible to market shocks, the possible effect of financialization on food security has become an evolving worry in recent years. Commodity market speculation can cause quick price increases, making it more difficult for low-income customers to obtain food. Furthermore, as investors prioritize investments in more significant, internationally linked markets, financialization may discourage investment in local and regional food systems.

Given these difficulties, more study on the effect of financialization on agricultural commodities and food security is required. Understanding the connection between farm financialization and financial disasters in global food security is critical for officials developing food security strategies. Examples of such tactics are increased investment in local and regional food systems, assistance for small-scale farms, and the creation of more transparent and equitable market mechanisms. This study seeks to add to a better comprehension of this complicated problem and influence policy choices to ensure a more stable and secure food system by analyzing the role of financialization in global food markets through commodities such as wheat, maize, and soybean. The remaining part of this study is structured as follows: Section 2 highlights the existing literature. Section 3 explains the econometric methodology used for the study and gives a description of the data used. Section 4 presents the empirical findings. Section 5 discusses the results and concludes in Section 6.

## 2. Literature review and hypothesis

Researchers have used the term 'financialization' to refer to a multitude of definitions. Historically, it has been considered to represent the increasing dominance of 'shareholder value' as a way to conduct corporate governance. A few researchers use the term to refer to the eminence of the capital market system as a financial system over traditional banking alternatives (Krippner, 2004; Manogna & Mishra, 2020, 2022b). The meaning of "financialization" that Epstein (2005) provided in the preface to his edited book "Financialization and the World Economy" is arguably the one that is most frequently referenced. He defines it as the process by which the functioning of the domestic and global economies becomes increasingly influenced by financial incentives, markets, actors, and organizations. In addition, Sawyer (2013) maintains that there are significant differences between the definitions mentioned by Epstein and Krippner. The first discusses the phases or eras of capitalism, whereas the second defines the field of study. The growing role of financialization in today's world is highly substantiated by data that shows that activity on financial markets has grown faster than actual activity; financial profits account for a more significant portion of total profits, and households and the financial sector are taking on a lot more debt (Stockhammer, 2010; Manogna & Mishra, 2021a). Research has shown that on a macroeconomic scale, the era of financialization has been linked to tepid economic growth, which shows a slowing trend (Palley & Palley, 2013; Manogna, 2021b; Manogna & Mishra, 2021b).

Given the different definitions authors have associated financialization with, multiple indicators have been developed to measure it. Kedrosky and Stangler (2011) compute it by dividing the size of the banking industry by GDP. Stockhammer (2010) employs non-financial companies' interest and dividend income as a surrogate for financialization. Freeman (2010) examines the financial sector's profit share, the ratio of financial-sector earnings to total private-sector wages and remuneration, and the ratio of financial assets split by GDP. Krippner (2005) compares financial and non-financial company earnings and portfolio income of non-financial companies. Assa (2012) introduces a composite variable utilizing two indicators: Financial value contributed as a percentage of overall value added and finance jobs as a percentage of overall employees. Commodity financialization refers to treating commodities, such as gold, oil, or agricultural products, as financial assets that can be traded and invested in, like stocks, bonds, or currencies. To measure the financialization of commodities, authors generally use an index associated with the commodity and its futures, such as the S&P Goldman Sachs Commodity Index (GSCI) (Adams & Glück, 2015; Manogana & Mishra, 2022a; Manogna & Mishra, 2023). Chari and Christiano (2017) introduces a measure of open interest calculated by measuring the volume of trade in the futures market with respect to global output. They also include a second measure called the net financial flow, which stands for the net position of speculators in relation to world output. For the sake of our analysis in this study, we construct a composite financialization variable involving the annual trading volume, annual open interest contracts, and a ratio of the annual trading volume to the annual open interest contracts for the respective commodities considered.

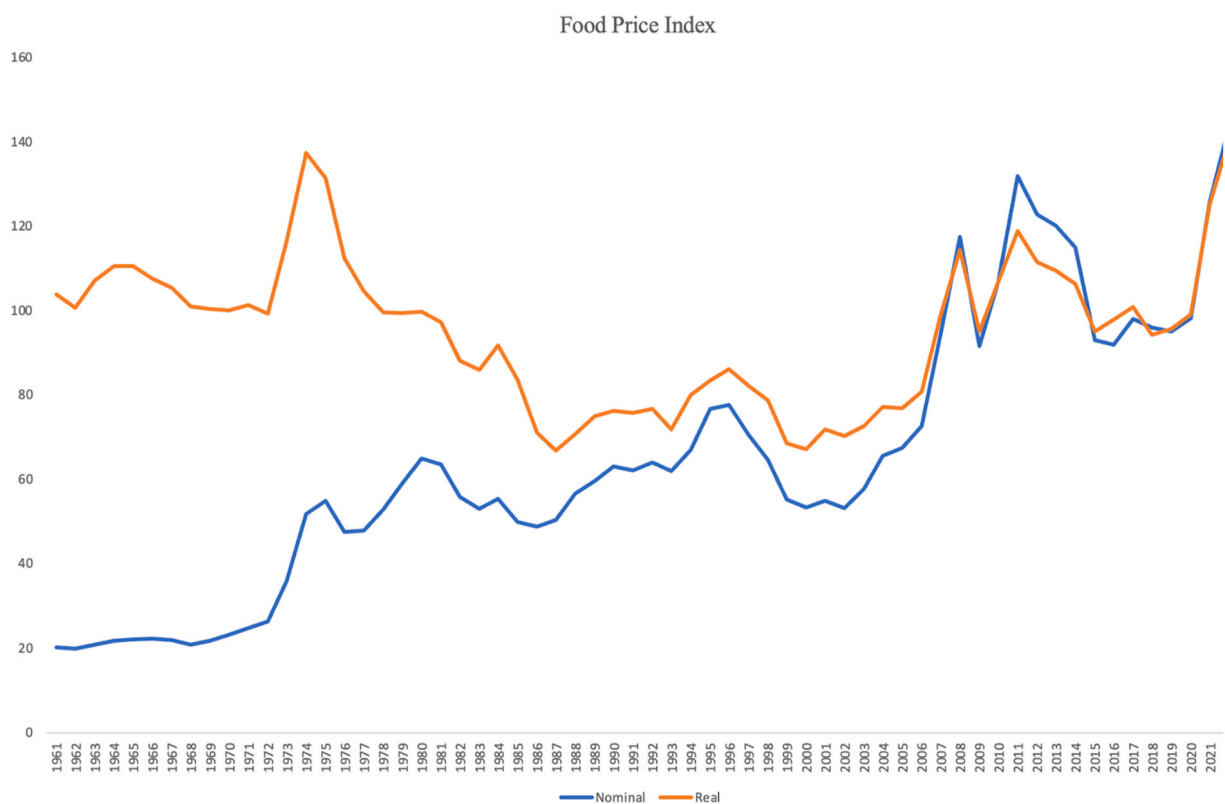

**Fig. 1.** Annual food price indices: 2014–16 = 100.
**Source** – Author's representation. UN FAOSTAT.

The financialization of agricultural products has been a topic of much discussion and research in recent years. One of the critical drivers of financialization has been the growing demand for investment opportunities in a low-interest-rate environment. This has been facilitated by the development of financial instruments such as commodity index funds and exchange-traded funds (ETFs), which enable investors to gain exposure to a diversified portfolio of commodities without worrying about the physical delivery of the underlying assets. The growth of financialization has been accompanied by a shift in the demeanor of market participants, particularly in the form of increased speculation and short-term trading. In the agricultural sector, using futures markets has enabled investors to take positions in agricultural commodities without connecting to the physical market. This has led to a greater focus on short-term price movements and reduced investment in long-term production and supply considerations. Increased financialization has been associated with increased price volatility in agricultural markets, as financial actors speculate on the prices of agricultural commodities. A study by Clapp and Helleiner (2012) found that financialization has contributed to increased price volatility in agricultural markets, which can lead to food insecurity. Tang and Xiong (2012) show that financialization increased the correlation between commodity prices and stock prices, suggesting that financialization may lead to increased speculation.

Similarly, Gilbert (2010) argues that using futures markets can lead to market distortions and exacerbate the impact of supply shocks, particularly in developing countries where farmers may not have access to the financial instruments necessary to manage price risk. Several hedge fund managers, commodity end-users, policymakers, and some economists contend that commodity index investment significantly drove the 2007–2008 spike in commodity futures prices (e.g., Baffes & Haniotis, 2010, p. 5371; Manogana & Mishra, 2022c). The massive wave of index fund buying allegedly created a "bubble" that forced commodity futures prices well above "fundamental values." This reasoning fuelled political pressure to limit speculative positions in commodity futures markets, particularly in energy futures markets. The Commodity Futures Trading Commission (CFTC) was then given the authority to set aggregate speculative position limits on futures and swap positions in all non-exempt "physical commodity markets" in the United States. Other studies have suggested that financialization can increase market concentration, negatively affecting small-scale farmers and food security. For example, a study by Cotula (2011) found that financialization can increase land grabbing, whereby large-scale investors acquire land in developing countries for agricultural purposes. This can displace small-scale farmers and negatively impact their livelihoods, which can, in turn, contribute to food insecurity. Youcef (2019) investigates the impact of financial investors on agricultural prices using the Threshold Autoregressive Quantile methodology and finds evidence of reinforcement linkages between equity and agricultural markets, showing that financial mechanisms have a greater impact on commodity markets during extreme movements, potentially influencing the financialization of commodities.

Another negative consequence of financialization is the promotion of monoculture farming. According to a study by Mamabolo et al. (2021), financialization promotes the production of a small set of staple crops, considered safer investments. This can lead to the neglect of other crops and reduce the diversity of agricultural production, which can negatively impact food security. A study by Headey (2011) found that the concentration of market power in the hands of a few prominent players could lead to distortions in market prices and supply chains, particularly in developing countries. The authors argued that this could adversely affect food security, reducing access to food for low-income consumers.

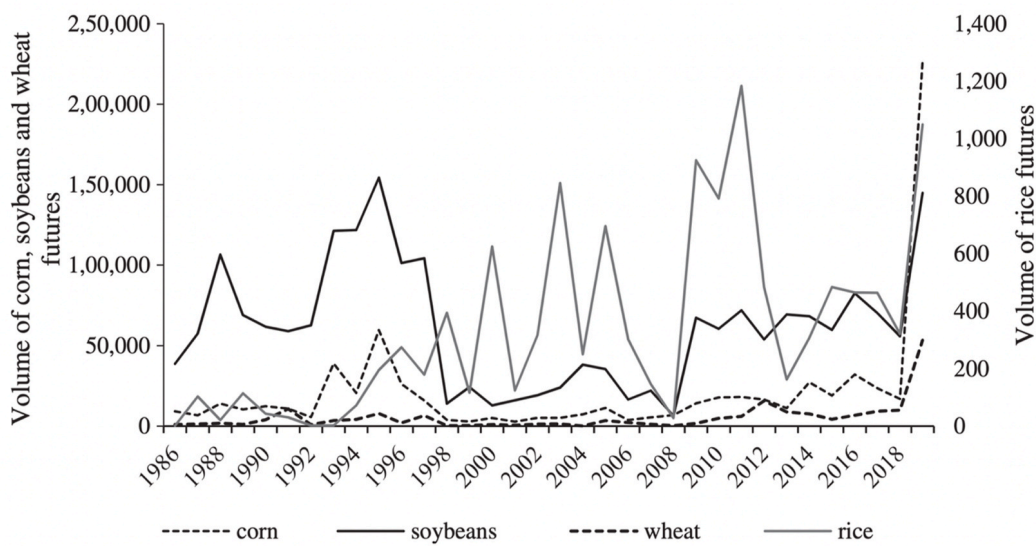

**Fig. 2.** The graph shows the fluctuations in commodity futures on corn, rice, soybeans, and wheat since the deregulation in the U.S. and elsewhere.
**Source** - Wiki Continuous Futures on Quandl database.

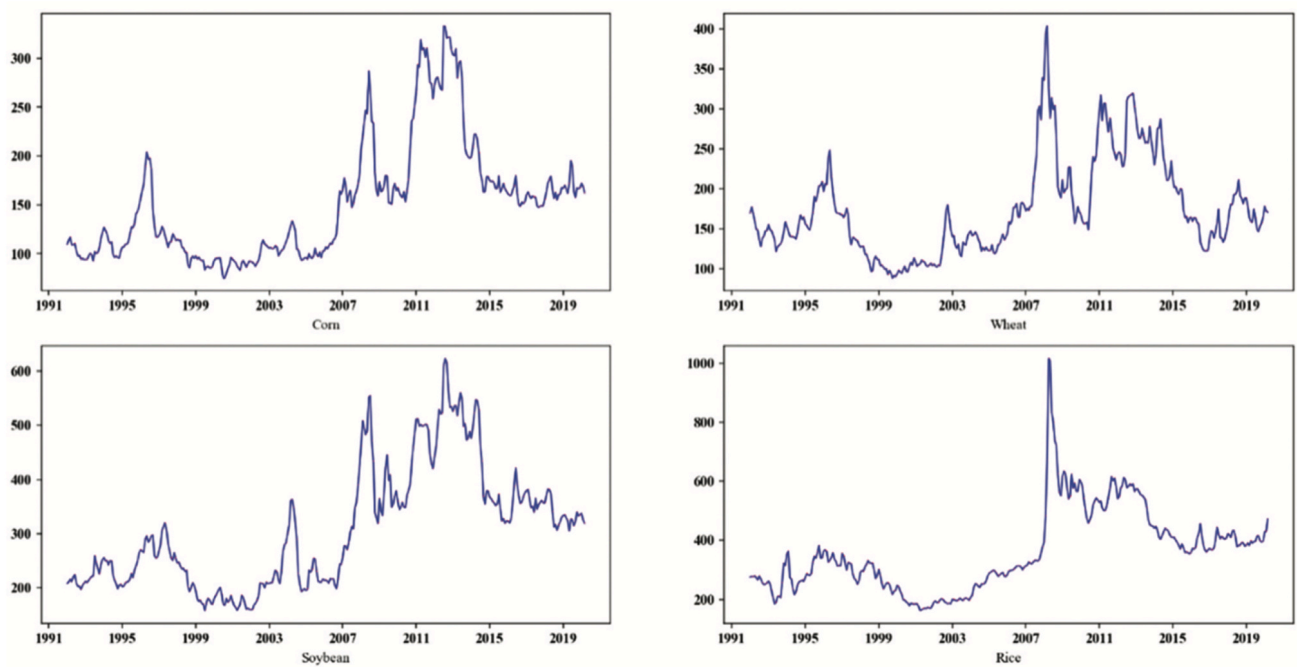

**Fig. 3.** The price fluctuations of the four major staple grains in the international market from 1992 to 2019. Unit: USD/ton.
**Source** - Author's representation. Following IMF, we select the FOB price of the U.S. No. 1 hard red winter wheat in the Gulf of Mexico, the U.S. No. 2 yellow corn in the Gulf of Mexico, 5% broken milled white rice from Thailand, and the No. 2 soybean futures contract price of the Chicago Board of Trade in the United States.

On a more positive note, some researchers argue that financialization can lead to greater market efficiency and risk management. For example, Irwin and Sanders (2011) present that many market analysts and economists have expressed skepticism about the bubble argument, citing logical inconsistencies and contrary facts. They stated that commodity markets in 2007–2008 were driven by fundamental supply and demand factors that pushed prices higher, not "excessive speculation." In the grain markets, the diversion of row crops to bio-fuel production and weather-related production shortfalls are cited, demand growth from developing nations, U.S. monetary policy, and a lack of investment in basic production infrastructure (e.g., Trostle, 2008; Wright, 2012; Manogana 2021a; Manogana & Mishra, 2021c). Irwin et al. (2009) present further analysis regarding the "bubble" hypothesis, showing that it does not hold true under close scrutiny. The authors also performed a causality test to show that positions in the commodity futures market do not consistently lead to future price changes. In more recent times, Etienne, Irwin, and Garcia (2018) used four different speculative measures (non-commercial net long positions, excessive speculative volume index, index trading activities, and Working's speculative index). They developed Structural Vector Autoregressive (SVAR) models to robustly state that the effect of speculation does not have an impact on the price of corn. Similarly, a study by Bellemare et al. (2013) found that financialization had no significant impact on the volatility of global food prices. A unique point of view is that financialization can benefit food security by increasing investment in agricultural production. A study by Clapp (2014) found that financialization can lead to increased investment in agricultural production, which can help to increase food supplies and improve food security. A study by Bohl et al. (2015) found that financialization positively impacted agricultural commodity prices, particularly in the case of corn and soybeans. However, the authors also noted that financialization could increase volatility, adversely affecting food security.

The link between financialization and food security has been brought into sharp focus by several food crises in recent years. The 2008 food crisis, which was accompanied by a spike in food prices and widespread protests and riots in many countries, is often cited as a key example of the potential impact of financialization on food security. In a study examining the causes of the crisis, Von Braun et al. (2012) applied Granger causality tests and found that the surge in the price of wheat, maize, soybean, and rice was partly driven by the growth of financial speculation in commodity markets. Sean Field (2016) presents an alternative take on global food price volatility compared to other studies that involve neoclassical literature. The author uses Marxian circuits of capital to identify the connection between food consumers and index swap dealers. Their findings show a positive relationship between financial speculation by hedge funds and swap dealers and food price volatility.

The potential impact of financialization on food security is a complex issue, and many factors may influence the relationship between the two. For example, the impact of financialization may vary depending on the type of commodity being traded, the degree of market concentration and regulation, and the level of economic development in the countries involved. Some studies have also suggested that financialization may have differential impacts on different groups within countries, such as rural farmers or urban consumers Clapp (2015). Lawson et al. (2021) consider four different food grains in their study. Their results show that the effect of speculation depends on the food grain considered. Traders' behavior was found to be significantly different when rice and wheat (which are typically used for human consumption) were compared to corn and soybean (a large percentage of which is animal feed).

Another approach researchers take is to evaluate the impact of speculation by considering different types of it. Bredin et al. (2021) consider two different forms of short-run trading consisting of a group called Manipulators (the biasing influencer in relation to the fundamental price) and the Speculators (correcting influence). The authors evaluate that the Manipulators play an increased role during periods generally associated with financialization. Another key driving factor behind the increasing food prices is described by multiple researchers as the energy price. Farhad et al. (2018), in their research across eight different Asian countries, conclude that energy prices significantly impact food prices. Their results show that the price of agricultural produce positively correlates to oil price shocks. That about 64% of the variance in food price is explained by movements in the oil price compared to other energy sources such as biofuel.

Researchers have blamed financialization for reduced food security in developed nations. Sosoo, Okorie, and Chen (2021) ran time series regression in lower, middle, and higher-income countries. They concluded that commodity futures have a more notable negative impact on food security in low-income countries than their counterparts. They also state that the financial crises have an impact on all the regions as a whole and only deepen the adverse effects that commodity futures present to food security. Tsui et al. (2017) show the impact of grain financialization on developing countries and China in particular. They establish that grain financialization is equivalent to U.S. Dollarization. Ouyang and Zhang (2020) study the effect of financialization on agricultural commodities in China. The researchers utilize a time-varying copula approach to study the stock markets' dependency on agricultural commodities. They identify a positive correlation between the two, and the correlation is time-varying. Given the potential risks associated with financialization, there have been calls for greater regulation and oversight of the commodity markets. Policymakers have advocated for position limits (which restrict the number of futures contracts that individual traders can hold) and greater transparency and disclosure requirements for market participants.

This study proposes the first hypothesis.

**H1a.** Higher the degree of financialization of agricultural produce, the more dramatic the volatility in their prices and the more significant the negative impact on food security.

The fact that market speculation and price volatility are related has been inherently debated. The Efficient Market Hypothesis states that the current asset price reflects all available information, so it is impossible to make consistent predictions of the future price using historical data. As explored by Irwin et al. (2009), there is no concrete evidence to show that market speculation can cause bubbles in commodity prices. On the other hand, research by Gilbert (2010) and several other scholars states the opposite. As more investors engage in financial speculation in agricultural commodity markets, staple foods such as rice, wheat, and maize prices have become more volatile (UNCTAD, 2013). Moreover, the impact of price volatility on food security is particularly acute for vulnerable populations who may already be struggling with poverty and food insecurity. Therefore, understanding the relationship between financialization, price volatility, and food security is crucial in ensuring a more stable and secure food system for vulnerable populations. In order to do so, this study proposes the aforementioned hypothesis.

This study proposes the second hypothesis.

**H1b.** Monetary Policy has a positive moderating effect on the impact of agricultural products on food security caused by the financialization of agricultural products.

Studies have discovered that monetary policy tools, such as interest and exchange rates, can significantly impact commodity prices and affect food security outcomes (Awokuse, 2010; Manogna et al., 2021). Additionally, research has shown that exchange rate policies can also affect food security outcomes, as they can influence the availability and affordability of food imports, which is crucial in countries that are net food importers (Jamora et al., 2010). Ghosh (2011) states that the financial deregulation in the United States resulted in increased speculative activity in the commodity markets that led to a dramatic rise in food prices during the 2008 crisis. In addition, it states that such crises exacerbate food insecurity by placing limits on fiscal policies and food imports in developing nations with balance-of-payments constraints. This results in a depreciation of the currency due to the exodus of capital, which has a negative impact on employment and limits the capacity of disadvantaged populations to purchase food. This study offers the aforementioned hypothesis in order to evaluate the influence of monetary policy in greater detail.

This study proposes the third hypothesis.

**H1c.** The financialization of agricultural commodities on food security has a threshold effect on economies at different development levels, with a more considerable impact on developing countries compared to developed countries.

Financialization has a more significant impact on developing countries than on developed countries due to their higher dependence on agriculture as a source of income and food security (UNCTAD, 2013). In developed countries, the impact of financialization on food security is relatively minor due to their more diverse economies and more developed financial markets, which can help absorb financialization's impact on agricultural commodities (Gorton & Rouwenhorst, 2006; Manogna & Aayush, 2023). There are very few studies that have investigated the effect of the financialization of agricultural commodities on food security worldwide as well as on developed and developing countries separately, until very recent times, especially at the individual agriculture commodity level. Developed countries often have more diverse agricultural systems and can leverage technological advancements to improve yields and productivity, reducing reliance on commodity markets (Gibbon & Ponte, 2005). In order to explore this further, this study introduces the aforementioned third hypothesis.

## 3. Data and methodology

### 3.1. Financialization index construction

The financialization of agricultural commodities is the most important explanatory variable in the investigation. For wheat, maize, and soybean futures traded on the Chicago Board of Trade, we utilize annual trading volume, annual open interest contracts, and a ratio of annual trading volume to annual open interest contracts. Yearly numbers for maize, wheat, and soybean futures are sourced from the Bloomberg database. To examine the effect of commodity financialization in influencing food security, we quantify the rise in speculative activity on the agricultural commodity market using these measures.

#### 3.1.1. Variable 1: annual trading volume of futures contracts

This indicates the annual volume of commodity futures traded on the Chicago Mercantile Exchange (CME). It entails combining multiple short-term futures contracts into a single long-term historical dataset. A greater trading volume shows that more traders are participating in these markets. This can also signify a huge number of short-term futures contracts (Robles et al., 2009).

#### 3.1.2. Variable 2: annual open interest in futures contracts

*Open interest* is the sum of all commodity futures contracts that have not yet been settled by delivery, exercise, or an opposite futures position. Open interest is produced when a trader enters a futures contract position. The position stays open interest until the trader establishes a counter-position or the contract expires (Robles et al., 2009). On the commodity futures market, a greater value for open interest may indicate a greater quantity of medium- and long-term futures contracts.

#### 3.1.3. Variable 3: ratio of annual trading volume to annual open interest in futures contracts

A rise or decrease in the ratio is anticipated to reflect speculative actions in the commodities futures market, assuming that most speculators choose to enter into short-term contracts as opposed to hedgers, who enter into long-term contracts to hedge against future price volatility. An increase in the number of short-term contracts executed by speculators will result in a rise in yearly trading volumes. Nonetheless, it will have little effect on the yearly open registered interest. This indicates a rise in this ratio. This ratio is therefore anticipated to also accurately reflect the activity of market speculators (Robles et al., 2009).

Since these three indicators differ in their measurement of the financialization of agricultural commodities, conclusions may be skewed if only one indicator is used. Consequently, this analysis uses the aforementioned three indicators as a foundation and principal component analysis to develop a composite index of agricultural commodity financialization. This index is calculated separately for wheat, corn, and soybeans, and by averaging the three, an overall agricultural commodity financialization index is constructed.

*3.2. Data description*

The research paper focuses on food security as the dependent variable. Food security is measured using indicators from the FAO database, which encompasses aspects such as price, income, accessibility, sufficiency, safety, and nutritional considerations. The data employed in this study is drawn from various reputable sources, including the FAO database, the World Bank's World Development Indicators, and the Bloomberg database. The study's scope narrows down to the analysis of wheat, corn, and soybean, as these commodities are highly traded on the Chicago Board of Trade and have global significance in both agricultural production and consumption. Rice is subject to various trading restrictions, tariffs, and international agreements due to its critical role as a staple food source. These factors can introduce complexities in analyzing the financialization-food security relationship. Thus, rice and other agricultural products are refrained from being selected as samples for the empirical analysis in this study. The choice of countries is based on a few key reasons. First, this paper selected countries that hold prominent positions in the global food supply, thereby influencing global food security. Additionally, this study considered countries with active trading in agricultural commodities, particularly within the derivatives markets. This approach allows us to explore the potential influence of financialization on food security. To ensure a balanced representation, we incorporated a mix of countries with varying levels of human development, as measured by the Human Development Index (HDI). By including both developed and developing economies, this study sought to account for a range of economic and social contexts that might interact with the financialization-food security dynamic. As a whole, this selection approach forms a robust and comprehensive basis for investigating the intricate relationship between financialization and food security across diverse economic and agricultural contexts. The data sample covers five developed countries - the United States, Australia, Canada, France, and Germany and seven developing countries - China, Russia, India, Indonesia, Brazil, Vietnam and Thailand. The research paper utilizes annual panel data spanning the period from 2000 to 2021. The primary explanatory factor under examination is the financialization of agricultural commodities, as elaborated earlier. Other control variables incorporated in the analysis consist of the annual GDP growth rate (expressed as a percentage), the annual consumer price inflation rate (expressed as a percentage), the proportion of arable land relative to total land area, the food price index (with the base year of 2015 set at 100), exchange rates (average for the period, expressed as local currency units per U.S. dollar), energy price index, food production index, and the natural logarithm of GDP per capita. Table 1 gives the definitions of all regression variables used in the study.

Within the scope of the twelve countries studied across the timeframe of 2000–2021, the food price index exhibits a range between 56.66 and 116.38, indicating significant variability in food prices. The composite financialization index (F.D.) reveals values spanning from −1.55 to 1.37. A graphical representation of the normalized global financialization index during 2000–2021 is depicted in Fig. 4. Notably, the global energy price index demonstrates a pronounced variance, suggesting substantial volatility in energy prices over the two-decade period. The food security variable (F.S.) spans from a minimum of 1.76 to a maximum of 14.78 across the sample. A comprehensive summary of the statistical properties of all variables is presented in Table 2. Furthermore, Fig. 5 portrays the food security index in developing economies throughout the 2000–2021 period, providing a visual depiction of its trends. Additionally, Table 3 presents a correlation matrix detailing the relationships between all variables utilized in the study, further enhancing our understanding of their interdependencies. The independent variables 'fpricein' and 'fprodin' are found to be highly correlated, hence we omit 'fpricein' from further analysis.

**Table 1**
Definitions of key regression variables.

| Variables | Definitions |
| --- | --- |
| F.S. | Food Security (Food security indicator from FAO considering price, income, accessibility, sufficiency, safety, and nutritional aspects of food security) |
| WF | Wheat Financialization is constructed by combining the indicator variables from 1 to 3 using the Principal Component Analysis for Wheat. |
| CF | Corn Financialization is constructed by combining the indicator variables from 1 to 3 using the Principal Component Analysis for Corn. |
| S.F. | Soybean Financialization is constructed by combining the indicator variables from 1 to 3 using the Principal Component Analysis for Soybean. |
| F.D. | The Composite Agricultural Financialization Index is constructed by combining the wheat, corn and soybean financialization using Principal Component Analysis. |
| *Control Variables* | |
| gdpgr | This variable represents the annual GDP growth rate. |
| arbl | This variable represents the total arable land of a country as a percentage of the total land area. |
| fpricein | This variable represents the annual Food Price Index. |
| infl | This variable represents the annual consumer price inflation. |
| energypri | This variable represents the annual Energy Price Index. |
| exchg | This variable represents the average annual exchange rate of local currency against the U.S. dollar. |
| fprodin | This variable represents the annual Food Production Index. |
| lngdppc | This variable represents the log of annual GDP per capita. |

Note: This table shows the definitions of the main dependent and independent variables used in the regression analysis in the study.

*3.3. Methodology*

*3.3.1. Basic panel data regression model*

This research paper employs a fixed-effects model to elucidate the impact of financialization on agricultural commodities on food security and to assess the varying degrees of this impact across developed and developing nations. The regression model is estimated as expressed in Equation (1):

$$FS_{i,t} = \alpha_i + \beta_0 + \beta_1 control_{i,t} + \beta_2 financialization_{i,t} + \mu_{i,t} \quad \ldots \ldots \quad \text{Eq 1}$$

Here, F.S. represents food security, which gauges the availability of food to fulfill daily nutritional needs at a time. The primary focal point in this study is financialization, and it is operationalized through W.F. (wheat futures), C.F. (corn futures), S.F. (soybean futures), and F.D. (composite financialization) using Principal Component Analysis (PCA). This technique generates distinct regressions for each commodity. The control variable encompasses all the factors used for control purposes within this investigation. The term μ at time t denotes random error, while the term α signifies the individual fixed-effects factor. Through Hausman's Test, the utilization of panel fixed effects regression models is justified over random effects, subsequently forming the basis of this study's analytical approach.

*3.3.2. Moderating effect model*

To examine the potential moderating influence of monetary policy on the relationship between the financialization of agricultural commodities and food security, we introduce an interaction term between these variables based on the framework outlined in the model (3.2.1).

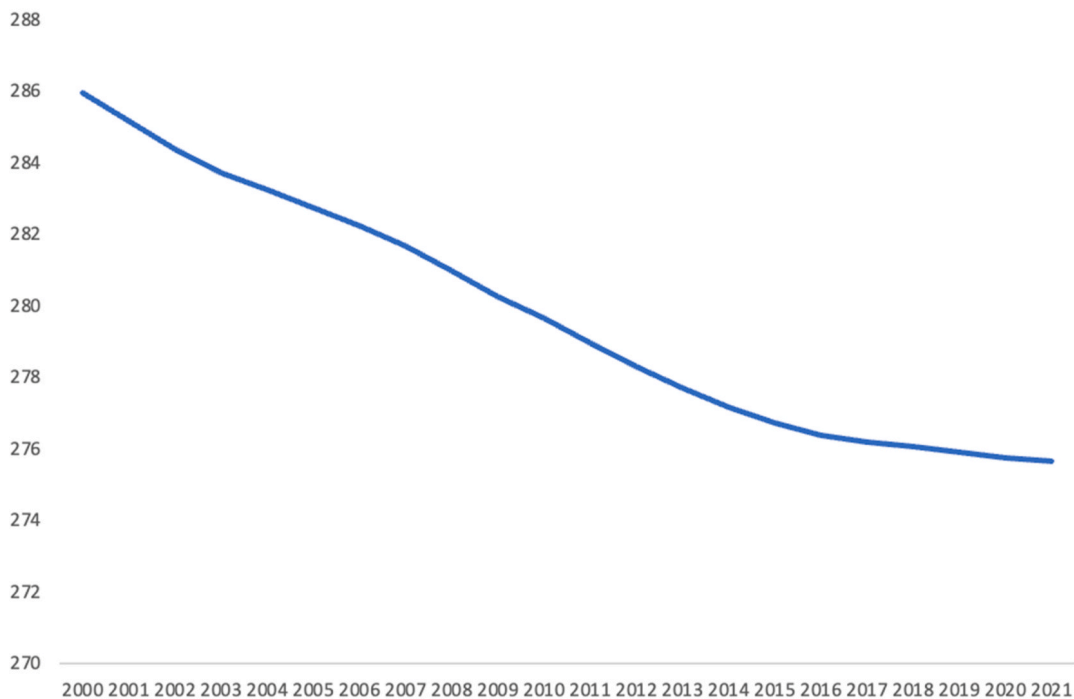

**Fig. 4.** The graph shows the average food security index in developing economies over the period 2000–2021.

**Table 2**
Summary statistics of key variables.

| Variable | Mean | Std. Dev | Min | Max |
|---|---|---|---|---|
| FS | 8.18 | 2.48 | 1.76 | 14.78 |
| WF | 0.00 | 0.98 | −1.46 | 1.52 |
| CF | 0.00 | 0.98 | −1.49 | 1.71 |
| SF | 0.00 | 0.98 | −1.52 | 1.34 |
| FD | 0.00 | 1.00 | −1.56 | 1.38 |
| energypri | 83.19 | 27.18 | 44.89 | 125.48 |
| fprodin | 80.44 | 26.11 | 15.53 | 143.12 |
| lngdppc | 9.74 | 0.91 | 7.64 | 11.15 |
| gdpgr | 3.67 | 3.36 | −7.81 | 14.23 |
| arbl | 19.77 | 14.67 | 3.01 | 54.13 |
| infl | 4.02 | 3.84 | −1.71 | 23.11 |
| exchg | 2502.13 | 5914.45 | 0.68 | 23208.37 |

This results in the formulation of the specific model presented as Equation (2).

$$FS_{i,t} = \alpha_i + \eta_0 + \eta_1 control_{i,t} + \eta_2 financialization_{i,t} \times M2 + \eta_3 M2 + \mu_{i,t} \ldots \ldots \text{Eq 2}$$

### 3.3.3. Panel regression using threshold variable

This research paper employs a panel threshold regression model to delve into the correlation between the financialization of agricultural commodities and food security across economies of varying developmental stages. The threshold variable chosen for this purpose is the Human Development Index (HDI). The regression model, as presented in Eqn (1), is estimated twice: once encompassing all observations where HDI is less than 0.85, and subsequently for observations where HDI is greater than or equal to 0.85. This bifurcation allows us to differentiate between the potential impact on food security in developing and developed nation-states, respectively.

$$FS_{i,t} = \alpha_i + \delta_0 + \delta_1 control_{i,t} + \delta_2 financialization_{i,t} + \mu_{i,t} \ (HDI<0.85) \ldots \ldots \text{Eq 3}$$

$$FS_{i,t} = \alpha_i + \delta_0 + \delta_1 control_{i,t} + \delta_2 financialization_{i,t} + \mu_{i,t} \ (HDI \geq 0.85) \ldots \ldots \text{Eq4}$$

## 4. Empirical results and analysis

### 4.1. Basic panel regression results

Table 4 presents the outcomes of the regression analysis covering the entire sample. Specifically, focusing on the influence of wheat futures trade on global food security, we observe a significant coefficient of −0.153 for wheat futures (W.F.) at a 1% level. This finding suggests that the trading of wheat futures has brought about a noteworthy negative impact on global food security. Similarly, when exploring the isolated effect of corn futures, the coefficient for corn futures (C.F.) stands at −0.038, signifying significance at a 1% level. In a similar vein, the coefficient for soybean futures (S.F.) is −0.1526, demonstrating significance at the 1% level. Furthermore, when assessing the collective effects of wheat, corn, and soybean, the coefficient for financialization (F.D.) is −0.1494, signifying statistical significance at a 1% level. These findings collectively underscore the pronounced influence of financialization on food security.

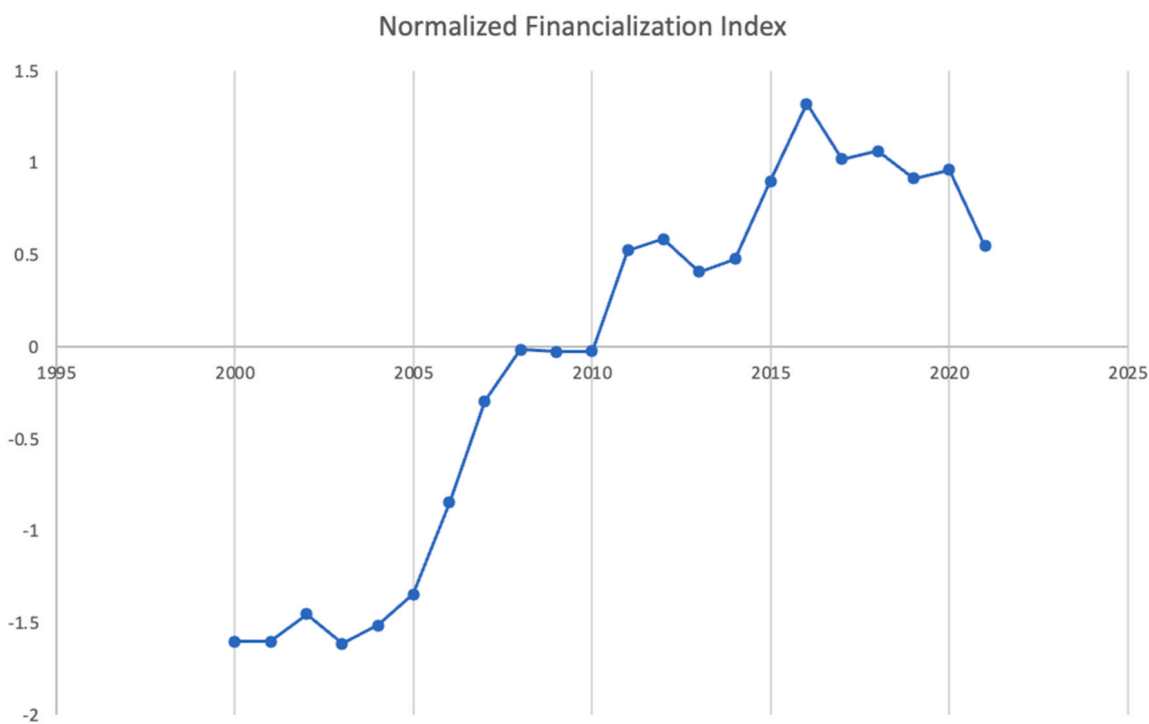

**Fig. 5.** The graph shows the normalized financialization (F.D.) index over the 2000–2021 period.

**Table 3**
Correlation Matrix of all variables used in the study.

|         | gdpgr   | arbl    | infl    | exchg   | energypri | fprodin | lngdppc | FD     |
|---------|---------|---------|---------|---------|-----------|---------|---------|--------|
| gdpgr   | 1.0000  |         |         |         |           |         |         |        |
| arbl    | 0.0501  | 1.0000  |         |         |           |         |         |        |
| infl    | 0.2236  | −0.1153 | 1.0000  |         |           |         |         |        |
| exchg   | 0.2471  | −0.0486 | 0.2039  | 1.0000  |           |         |         |        |
| energypri | 0.1127 | −0.0031 | 0.1176 | 0.0069  | 1.0000    |         |         |        |
| fprodin | −0.4008 | 0.0439  | −0.2628 | −0.0738 | 0.1878    | 1.0000  |         |        |
| lngdppc | −0.3512 | −0.2844 | −0.3715 | −0.3822 | 0.1383    | 0.3680  | 1.0000  |        |
| FD      | −0.2390 | 0.0086  | −0.1708 | 0.0646  | 0.3857    | 0.3869  | 0.3254  | 1.0000 |

The outcomes of these regression analyses highlight a consistent trend: the financialization, particularly futures trading, of all three key agricultural commodities - wheat, corn, and soybean - has yielded a detrimental impact on global food security. When scrutinized individually, wheat and soybean exhibit more pronounced negative impacts on food security compared to corn, as evidenced by the magnitude of their respective coefficients. When considered collectively, the results affirm that the overall financialization of agricultural commodities significantly and detrimentally affects food security, with magnitudes akin to those observed when wheat and corn are evaluated in isolation. The extensive degree of financialization within the realm of agricultural produce also serves as an indicator of heightened volatility in food prices.

In terms of other control variables, the coefficients hold notable insights. The energy price index, significant at a 1% level, reveal smaller magnitudes that contribute less significantly to variations in global food security. The annual consumer price inflation, significant at a 5% level, and the exchange rate, significant at 1%, also exhibit relatively smaller influences. Conversely, the food production index, significant at a 1% level, exerts a substantial positive influence on global food security.

Furthermore, the estimation outcomes suggest that the cumulative effect of external explanatory factors can account for a substantial portion, at least 60 percent, of the observed variability in food security. Importantly, the joint statistical significance of these explanatory factors is confirmed by the F-Statistics tests, which are significant at the 1% level. This collectively underscores the significance of these factors in shaping global food security dynamics.

*4.2. Robustness test results*

In order to check the robustness of our findings, this study draws on

*M. R. L. and N. Kulkarni*  Borsa Istanbul Review 24 (2024) 280–291

**Table 4**
Effects of agricultural commodity futures on global food security.

|  | Food Security | | | |
|---|---|---|---|---|
|  | Only Wheat Futures (W.F.) | Only Corn Futures (C.F.) | Only Soybean Futures (S.F.) | Combined Futures (F.D.) |
| W.F. | −0.1531***<br>(-0.0197) | – | – | – |
| C.F. | – | −0.0382***<br>(0.0130) | – | – |
| S.F. | – | – | −0.1526***<br>(-0.0203) | – |
| F.D. | – | – | – | −0.1494***<br>(-0.0199) |
| gdpgr | 0.0032<br>(0.0038) | 0.0013<br>(0.0041) | −0.0005<br>(0.0038) | 0.0002<br>(0.0040) |
| arbl | −0.0219<br>(0.0150) | −0.0240<br>(0.0163) | −0.0211<br>(-0.0146) | −0.0228<br>(-0.0146) |
| infl | 0.0073**<br>(0.0034) | 0.0078**<br>(0.0038) | 0.0079**<br>(0.0035) | 0.0073**<br>(0.0036) |
| energypri | 0.0012***<br>(0.0035) | 0.0011***<br>(0.0004) | 0.0013***<br>(0.0004) | 0.0013***<br>(0.0004) |
| exchg | 0.0000***<br>(0.0000) | 0.0000***<br>(0.0000) | 0.0000***<br>(0.0000) | 0.0000***<br>(0.0000) |
| fprodin | 0.1078***<br>(0.0010) | 0.1055***<br>(0.0010) | 0.1079***<br>(0.0012) | 0.1069***<br>(0.0010) |
| lngdppc | −0.4930***<br>(0.0680) | −0.6905***<br>(0.0662) | −0.5217***<br>(0.0655) | −0.5509***<br>(0.0698) |
| Constant | 4.1120***<br>(0.6989) | 6.1685***<br>(0.6979) | 4.2826***<br>(0.6976) | 4.2826***<br>(0.6976) |
| Overall $R^2$ | 0.7274 | 0.7142 | 0.7091 | 0.6452 |

Notes: Robust standard errors are reported in parentheses. *$p < 0.1$; **$p < 0.05$; ***$p < 0.01$.

the generalized method of moments estimation (GMM) proposed by Blundell and Bond (1998) to conduct robustness tests on the model. Table 5 reports the regression results of GMM estimation. The coefficient of the core explanatory variable of interest in this study, financialization of agricultural commodities (F.D.), is positive, which is significant at the 9% level in the dynamic panel data model, and the results are consistent with the baseline regression results, indicating that the results obtained in this study are robust.

*4.3. Food security impact on developing versus developed nations*

To examine hypothesis H1c and investigate the distinct effects of agricultural commodity financialization across economies of varying developmental stages, this study adopts the Human Development Index (HDI) as the differentiating criterion. Specifically, the HDI value of 0.85 is utilized to categorize economies into developed (HDI ≥0.85) and developing (HDI <0.85) countries. The HDI serves as a composite metric of human well-being, amalgamating indices for health, education, and income.

Detailed regression results for the entire sample are provided in Table 6, with separate analyses conducted for developed and developing economies. The overall agricultural commodity financialization index, previously established, is employed for these analyses.

In developed economies, the coefficient for financialization (F.D.) does not exhibit statistical significance. However, in the case of developing economies, the coefficient stands at −0.177, signifying significance at a 1% level. This underscores that the financialization of agricultural commodities bears a substantial and negative impact on food security within developing economies. This could potentially be attributed to the fact that developing economies are in the process of

**Table 5**
The results of GMM robustness test estimates.

|  | Coefficient | Panel-corrected standard error | Z-statistic |
|---|---|---|---|
| L.FD | 0.1918* | 0.1112 | 1.7300 |
| L.fs | −0.4724*** | 0.1780 | −2.6500 |
| L.gdpgr | −0.0102** | 0.0052 | −1.9800 |
| L.infl | −0.0248* | 0.0134 | −1.8600 |
| FD | 0.1400*** | 0.0486 | 2.8800 |
| arbl | 1.0930*** | 0.3531 | 3.0900 |
| infl | 0.0134*** | 0.0037 | 3.6100 |
| energypri | 0.0043*** | 0.0015 | 2.8100 |
| exchg | 0.0000**** | 0.0000 | 2.8400 |
| fprodin | 0.0990*** | 0.0032 | 31.0800 |
| lngdppc | −2.2320*** | 0.7193 | −3.1000 |
| Constant | 0 | (omitted) |  |

Note: *$p < 0.1$; **$p < 0.05$; ***$p < 0.01$.

**Table 6**
The results of threshold regression with HDI.

|  | Food Security | |
|---|---|---|
|  | Developed nations | Developing nations |
| FD | −0.0014<br>(0.0043) | −0.1772***<br>(0.0423) |
| gdpgr | −0.0007<br>(-0.0006) | −0.0041<br>(-0.0040) |
| arbl | 0.0076**<br>(-0.0030) | 0.0962***<br>(-0.0164) |
| infl | −0.0056***<br>(.0015) | −0.0019<br>(0.0033) |
| energypri | 0.0003***<br>(0.0000) | 0.0009*<br>(0.0005) |
| exchg | 0.0522***<br>(0.1210) | 0.0000***<br>(0.0000) |
| fprodin | 0.0942***<br>(0.0004) | 0.0079***<br>(0.0024) |
| lngdppc | −0.8195***<br>(0.0217) | −0.7997***<br>(0.0684) |
| Constant | 8.5642***<br>(0.2262) | 8.5268***<br>(0.8115) |
| Overall $R^2$ | 0.7896 | 0.7407 |
| F Statistic | 101512.57*** | 11462.01*** |

Notes: Robust standard errors are reported in parentheses. *$p < 0.1$; **$p < 0.05$; ***$p < 0.01$.

establishing robustly regulated commodity futures markets. Consequently, these markets may experience higher influxes of speculators, resulting in elevated agricultural price bubbles. This, in turn, renders these economies more susceptible to agricultural price volatility, thereby jeopardizing their food security.

*4.4. Regression results of moderating effects*

Presented in Table 7 are the outcomes of the regression analysis exploring the moderating impact of monetary policy. With the inclusion of the interaction term involving the financialization of agricultural commodities and monetary policy, noteworthy insights emerge. The coefficient associated with the interaction term (M2) presents a positive and statistically significant trend at the 1% level. This observation signifies a positive moderating influence of monetary policy on the nexus between the financialization of agricultural commodities and food security. Moreover, when examining the coefficient resulting from the multiplication of financialization and M2, a negative value is observed, holding statistical significance at the 1% level; this finding underscores that financialization has introduced an adverse impact on food security. The massive input of liquid money brought by the quantitative easing of monetary policy further stimulates speculators' investment in tangible commodities. The massive influx of financial capital into the commodity futures market further stimulates the sharp rise in food prices and exacerbates food security risks.

## 5. Results discussion

The regression results show that the coefficients of wheat, corn, and soybean futures are negative and significant. A negative and significant coefficient is also obtained when considering all three commodities as an aggregate. The coefficients acquired for the other control variables are smaller in magnitude and do not make significant contributions to the volatility in global food security. A robustness check using the General Method of Moments estimation shows consistent results compared to the base regression models, establishing its robustness. These results confirm our initial hypothesis that the higher the degree of financialization of agricultural produce, the more dramatic the volatility in their prices and the more significant the negative impact on food security. A threshold regression was conducted to identify if there are significant differences in the impact that financialization has on developed and developing economies. It shows strong results that the financialization of agricultural commodities significantly impacts food security in developing economies. Using a Human Development Index (HDI) value of 0.85 as the horizon between the developed and developing economies, we obtain a positive coefficient that is statistically significant for developing countries. These results help confirm our third hypothesis that the financialization of agricultural commodities on food security has a threshold effect on economies at different development levels, with a more considerable impact on developing countries than developed countries. The regression results obtained after adding another variable to signify the interaction between the financialization of agricultural products and monetary policy produce a positive and significant coefficient. This shows that the injection of liquid money by easing monetary policy stimulates speculators' investment in tangible commodities. The massive influx of financial capital into the commodity futures market fuels the sharp rise in food prices. It exacerbates food security risks, which agrees with our second hypothesis that monetary policy has a positive moderating effect on the impact of agricultural products on food security caused by the financialization of agricultural products.

Understanding the relationship between the financialization of agricultural products and financial crises in global food security is crucial for policymakers to develop strategies for promoting food security. Additionally, the results obtained starkly contradict those established by Bellemare et al. (2013), who stated that financialization had no significant impact on the volatility of global food prices. A study by Bohl et al. (2015) found that financialization positively impacted agricultural commodity prices. Our research shows that introducing particular variables can help obtain more robust results.

## 6. Conclusion and implications

The worldwide food security scenario is dire, and each nation is attempting to guarantee its food security by lowering food prices and increasing food sovereignty. Meanwhile, as agricultural goods become more financialized, the impact of macroeconomic and financial markets on food security becomes more pronounced. Each country's food security risk is growing as a result of the present global economic downturn and greater volatility in financial markets. This article uses the contracts for wheat, corn, and soybean traded on the Chicago Board of Trade to analyze the role of commodity financialization in influencing food security and arrive at the following conclusions.

Firstly, the basic panel regression analysis shows that the financialization of all three commodities (wheat, corn, and soybean) has negatively affected their food security index by up to a significance of 1%. Additionally, a robustness test conducted using the Generalized Method of Moments (GMM) estimation produces results consistent with the baseline regression results, reinforcing its robustness.

Secondly, this article presents a panel threshold model to study the effects of agricultural financialization on food security in countries at different stages of growth. The results show that the coefficient of financialization (F.D.) is insignificant in the case of developed economies. It has a substantial threshold influence on food security, with a more significant negative impact in emerging countries. Because of an increasingly centralized global food and farming system, emerging countries have low food self-sufficiency, a rising import dependency, and a greater reliance on foreign aid. At the same time, because of their weak financial systems and inadequate agricultural financial regulating systems, emerging countries face more considerable food security risks than developed countries.

Additionally, this paper examines monetary policy's moderating influence on the impact of agricultural financialization on food security, finding that monetary policy has the potential to reduce the impact of agricultural financialization on food security positively. Lax monetary policy can generate excess market money, increasing farming speculation and aggravating food security risks. The financial sector should improve macroprudential monetary policy management and farm

**Table 7**
Moderating affect test results.

|  | Food Security | | |
| --- | --- | --- | --- |
|  | Coefficient | Std. Error | T-statistic |
| FD x M2 | −0.0021*** | 0.0003 | −6.45 |
| M2 | 0.0014*** | 0.00043 | 3.15 |
| *gdpgr* | −0.0041 | 0.0042 | −0.97 |
| *arbl* | −0.0911*** | 0.0281 | −3.25 |
| *infl* | 0.0076* | 0.0049 | 1.60 |
| *energypri* | −0.0009* | 0.0005 | −1.94 |
| *exchg* | 0.0017 | 0.0021 | 0.82 |
| *fprodin* | 0.1031*** | 0.0015 | 69.82 |
| *lngdppc* | −0.5886*** | 0.0700 | −8.40 |
| Constant | 6.6839*** | 0.9394 | 7.11 |
| F Statistic | 6.94*** | | |

Notes: *p < 0.1; **p < 0.05; ***p < 0.01.

derivatives market oversight to mitigate the detrimental effects of monetary policy changes on food security.

Given these results, it is evident that farm product financialization requires a balanced strategy. Governments and officials should encourage responsible investment in the sector while safeguarding small-scale farmers and local communities against the negative impacts of financialization. Land tenure changes, better access to loans and other financial services for small-scale farmers, and rules to prevent excessive trading and price fluctuations in agricultural markets are some of the answers. It is also important to remember that financialization is not the only factor affecting food security. Climate change, population expansion, and shifting nutritional patterns play a role, and any attempts to combat food poverty must take a holistic strategy considering all of these variables.

The results shed light on the complicated connection between financialization and food security. While financialization can have both positive and negative repercussions, it is evident that prudent investment and regulation are required to ensure that the benefits of financialization are realized while any negative consequences are mitigated. Developing nations should develop a food security policy based on self-sufficiency, peasant and organic agriculture, rural–urban collaboration, and improved food storage capacity. In order to defend themselves against international speculative money, they should also enhance the oversight and regulation of futures and financial markets. A balanced and comprehensive strategy can assist policymakers in ensuring that the world's food supply stays secure and available. However, caution must be taken not to overregulate the commodities market, as this could diminish market liquidity and impede the price discovery and risk management functions of futures markets.

The agricultural futures of other commodities have yet to be included in the analysis due to the unavailability of data. Future studies can incorporate proxy variables to conduct a more in-depth analysis of the impact of financialization. Additionally, exploring the impact of recent events, such as the Russia-Ukraine conflict and the food security issues in Pakistan, can provide a more holistic view. It is essential to recognize that climate change has a significant impact on the world's food supply. However, the scope of this study is limited to the function of financial derivatives. Future research may be conducted to investigate how climate change influences the findings.

Dr. Manogna R L is currently Assistant Professor in Economics and Finance at BITS Pilani, K.K. Birla Goa Campus. Her interest includes agricultural commodity markets, agricultural productivity, financial econometrics and firm performance